\documentclass[letterpaper]{article}
\usepackage{aaai19}
\usepackage{times}
\usepackage{helvet}
\usepackage{courier}
\usepackage{color}
\usepackage{makeidx}  
\usepackage{multirow}      
\usepackage{subfigure,graphicx}
\usepackage{epstopdf}
\usepackage{subfig}
\usepackage{amsmath,dsfont}
\usepackage{amsfonts}
\usepackage{graphicx,epstopdf}
\usepackage{enumitem}
\usepackage{color}
\usepackage{url}
\usepackage{bbding}
\usepackage{wasysym}
\usepackage[T1]{fontenc}
\usepackage{colonequals}
\usepackage{psfrag}
\usepackage{subfigure}
\usepackage{mathrsfs}
\usepackage{lscape}
\usepackage{setspace}
\usepackage{pifont,textcomp,bbm}
\usepackage{algorithm, algpseudocode}
\usepackage{dsfont}
\usepackage{amsmath,calc}
\usepackage{multicol}
\usepackage{booktabs}

\newtheorem{proof*}{Proof}
\newcommand{\nonl}{\renewcommand{\nl}{\let\nl\oldnl}}

\begin{document}
\title{Next Hit Predictor - Self-exciting Risk Modeling for Predicting Next Locations of Serial Crimes}

%
\author{Yunyi Li  \and \textbf{Tong Wang} \\University of Iowa \\ \{yunyi-li,tong-wang\}@uiowa.edu}

%
\maketitle              
\begin{abstract}
Our goal is to predict the location of the next crime in a crime series, based on the identified previous offenses in the series.  
We build a predictive model called \emph{Next Hit Predictor} (NHP) that finds the most likely location of the next serial crime via a carefully designed risk model. The risk model follows the paradigm of a self-exciting point process which consists of a background crime risk and triggered risks stimulated by previous offenses in the series. Thus, NHP creates a risk map for a crime series at hand. To train the risk model, we formulate a convex learning objective that considers pairwise rankings of locations and use stochastic gradient descent to learn the optimal parameters. Next Hit Predictor incorporates both spatial-temporal features and geographical characteristics of prior crime locations in the series. Next Hit Predictor has demonstrated promising results on decades' worth of serial crime data collected by the Crime Analysis Unit of the Cambridge Police Department in Massachusetts, USA.
\end{abstract}
\section{Introduction}
Crime prediction is an imperative step in predictive modeling. 
Previous works in the literature proposed different solutions to predicting the locations for future crimes \cite{wang2012automatic,bogomolov2014once,mohler2014marked,kang2017prediction}. 
Despite the increasing body of work in crime prediction, there has been a limited effort in predicting locations for \emph{serial crimes}. 
 Predicting a serial crime is different from a general crime prediction problem. Generic crime predictions often focus on modeling the distributions of crimes using historical events \cite{chainey2008utility,mohler2014marked}, for example, predicting ``hotspots'' which
is spatially localized area where many crimes occur. But these methods cannot differentiate the offenders of crimes and only identify spatial-temporal patterns that apply to the entire criminal cohort. Serial crime prediction, on the other hand, needs to consider the specific preferences of individual criminals. This is because different offenders may demonstrate different behavioral patterns when committing crimes \cite{gibson2013preferences,salo2013using}. For example, some criminals like to hit wealthy neighborhoods while some may choose to go to the more suburban area with houses more isolated from each other.  A serial crime can also happen in unusual locations where most of the other criminals are not interested in and therefore cannot be captured by generic crime prediction methods. This poses a significant modeling challenge compared to general crime prediction.  


Here, we propose a novel machine learning model, \emph{Next Hit Predictor}, to predict the location of the next serial crime committed by an unknown criminal, given his/her previous offenses and all crime events that happen in this area of interest. (``Unknown'' means we do not know any information about the criminal but only know the crimes committed by him/her.) 
We carefully design a self-exciting risk model that evaluates how likely the next crime is going to happen in each location (grid cell) in a map. The risk model consists of a \emph{background risk}, which is the average crime rate over a long period, and a \emph{triggered risk}, which is the sum of a set of kernels triggered by each of the prior offenses of the specific criminal of interest. The background risk represents the criminals' general preference for crime locations, and the triggered risk is designed to account for a criminal's individual preference for particular types of locations, which is motivated by the theory that criminals tend to operate in similar crime scenes since they feel familiar and safer \cite{warren1998crime}.  The rational criminal theory assumes that individuals have specific reasons for committing a crime at a certain time and a certain location \cite{clarke1980situational}.   By examining the historical crime instances committed by a serial criminal we can discover series-specific patterns that indicate criminals' preferences for executing crimes.

We design the triggering risk function as a fully parameterized isotopic kernel that measures the similarity between a location of interest and a previous crime location, considering spatial-temporal distances and the geographic features. The geographic features include urban and environmental factors, for example, for each location, we compute the area of residential buildings, the area of open space, the number of bus stops or subway stations, etc. 

To rigorously train the parameterized risk model, we
propose a learning objective built from comparing true locations of with other locations in the map for every crime in every series. The loss function for each pair of ranking is represented with a hinge loss to form a convex objective and an $l_2$ norm for regularization. 
 Then we apply a stochastic gradient descent algorithm to optimize the objective.

\section{Next Hit Predictor Model} \label{sec:model}
We investigate a region of interest $\mathcal{G}$ (the city of Cambridge in MA, USA) and a set of $n$ crime instances $\mathcal{C} = \{c_i, p_i\}^n_{i=1}$ located in this region. $c_i$ represents crime instance $i$ and $p_i \in \mathbb{Z}^P$ is its label. $p_i = 0$ means crime $i$ is not a serial crime but a singleton offense and $p_i>0$ represents the crime $i$ is in the $p_i$-th crime series. Let $\mathcal{C}_p$ represent indices of crimes in series $p$, i.e., $\mathcal{C}_p  = \{i|p_i = p\}$. 
 Each crime instance $c_i$ is described by three types of features, $c_i = \{s_i, t_i, \boldsymbol{\omega}_i\}$, where $s_i$ represents the spatial feature, i.e., the latitude and longitude of crime $i$,  $t_i$ represents the time when the crime happened, and $\boldsymbol{\omega}_i \in \mathbb{R}^d$ is a set of environmental and urban features associated with the grid cell crime $i$ was in.  Following the common practice in spatial-temporal machine learning \cite{lin2018grid,zhang2016dnn}, we discretize $\mathcal{G}$ into a grid map with a set of grid cells $\mathcal{L}$, indexed by $l$.  We use $g_i$ to represent the grid cell that crime $i$ is located in and $g_i \in \mathcal{L}$.

\subsection{Risk Model}
We formulate a model $r^{\langle p, t\rangle}_l$ to represent the risk of the next serial crime happening in a grid cell $l$, evaluated at time $t$, given the set of crimes in $\mathcal{C}_p$ that are committed by the same offender before time $t$. The proposed model follows the structure of a self-exciting point process that the risk consists of a background risk and triggered risks stimulated by previous offenses. The triggering function combines spatial, temporal and geographic features. The risk model is:
\begin{equation}
r^{\langle p, t\rangle}_l= \mu_l^{\langle t \rangle} + \sum_{i \in \mathcal{C}_p, t_i<t}\kappa(s - s_i, t - t_i, \boldsymbol{\omega}_l - \boldsymbol{\omega}_{g_i}),
\end{equation}
Here $\mu_l^{\langle t \rangle}$ indicates the \emph{background risk}, which represents how likely any crime will happen at location (grid cell) $l$, regardless of the preferences of any specific criminal. Therefore, $\mu_l^{\langle t \rangle}$ is independent of crime series and it captures the common preferences that apply to the entire criminal cohort. Here we choose $\mu_l^{\langle t\rangle}$ to be estimated from the most recent two years using kernel density estimation. 
 $\kappa(s - s_i, t_s - t_i, \boldsymbol{\omega}_l - \boldsymbol{\omega}_{g_i})$ is a kernel that represents the risk triggered by previous crime $i$ in the series $\mathcal{C}_p$. It measures the similarity between grid cell $l$ with grid cell $g_i$ (the grid cell where crime $i$ happened in) considering the spatial, temporal and geographic features. The sum of the kernels represents the risk triggered by all previous crimes in the series $\mathcal{C}_p$. Locations that are similar to the previous crime locations have a higher chance to attract the criminal. Compared to the background risk, the triggering kernels capture the individual preferences of an unknown criminal inferred from the previous offenses, which is independent of the general preference of the entire criminal cohort. Therefore, the risk model $r^{\langle p, t\rangle}_l$ combines criminals' general preferences for crime locations and criminal specific preferences. Using formula (1) we can create a risk map covering all grid cells in $\mathcal{L}$. 

Various forms of $\kappa(\cdot)$ have been proposed in the literature with the magnitude of the risk decreasing in space and time away from each previous crime \cite{yang2013mixture,zhao2015seismic,lewis2011nonparametric}, but most of the previous models only consider the effect of space and time. Here we design an isotropic kernel that incorporates the geographic similarity between each pair of grid cells, using urban and environmental features associated with each grid cell.  The triggering kernel is
\begin{equation}
\kappa(\Delta s,\Delta t,\Delta \boldsymbol{\omega}) = \frac{\beta_0+\sum_{j}\beta_j \Delta w_j}{(\Delta t+c)^2(\Delta s+d)^2},
\end{equation}
where $\Theta = \{c,d,\beta_0, \cdots, \beta_J\}$. $\{\beta_0, \cdots, \beta_J\}$ represents the weights for each urban and environmental features in evaluating the similarity and $\beta_0$ is an intercept. $\Delta s,\Delta t,\Delta \boldsymbol{\omega}$ represent the difference in spatial, temporal and geographic (urban and environmental features). $\Delta s$ is the Euclidean distance between two points. $\Delta t$ is the difference of the day that a previous crime happens and $t$. Since our model does not aim to predict the time of the next hit so we always choose $t$ to be one day after the last previous crime in $\mathcal{C}_p$ happens.

\subsection{Learning Objective}
We formulate a rigorous training objective to find the right parameters such that the true grid cell where a crime happens gets the highest risk. 
Let $l^{\langle p, t\rangle}_*$ represent the true grid cell for crime series $\mathcal{C}_p$. Given the risk model above, our goal is to find appropriate parameters $\Theta$ such that given a crime series $p$, $\forall l \in\mathcal{L}$, $r^{\langle p, t\rangle}_{l^{\langle p, t\rangle}_*} > r_l^{\langle p, t\rangle}.$
We will write $r^{\langle p, t\rangle}_{l^{\langle p, t\rangle}_*}$ as $r^{\langle p, t\rangle}_{*}$ for simpler notation. 
We formulate the parameter learning as a ranking problem. 
We define the loss function for each tuple $(p, t, l)$ as the
\begin{equation}\label{eqn:hinge}
\Lambda^{\langle p, t \rangle}(l) = \max\{0,r_l^{\langle p, t\rangle}-r^{\langle p, t\rangle}_{*}\}.
\end{equation}
This loss function means if $r_l^{\langle p, t\rangle}<r^{\langle p, t\rangle}_{*}$, then $l_*^{\langle p,t \rangle}$ is ranked before location $l$, so there is no loss. Otherwise the loss is the minimum necessary change for $r^{\langle p, t\rangle}_{*}$ in order to be correctly ranked before location $l$.  
We place an $l_2$ norm on $\beta$'s and the learning objective over all crimes in all series is 
\begin{equation}\label{eqn:newobj}
\Lambda = \sum_{p \in \{1,\cdots,P\}}\sum_{i \in \mathcal{C}_p}\sum_{l\in \mathcal{L}, l\neq l^*_p}\max\{0,r_l^{\langle p, t\rangle}-r^{\langle p, t\rangle}_{*}\} + \lambda_\beta ||\beta||^2
\end{equation}
Our goal is to find the optimal parameters $\Theta^*$ that
$
\Theta^*= \arg\min_{\Theta}\Lambda.
$

Directly optimizing (\ref{eqn:newobj}) is time consuming since it involves comparing risks for $|\mathcal{L}|-1$ location with the true location for every crime in every crime series. Therefore, we use a stochastic gradient descent algorithm on the bootstrap samples: we first draw a crime series $p$, and then randomly draw a crime $i$  and a location $l$ to compare with the true location that crime $i$ happened in. 
Then we apply stochastic subgradient with momentum to learn the parameters $\Theta$.

\subsection{Constructing Geographic Features}\label{sec:geographic}
In this section, we describe how to construct geographic features $\boldsymbol{\omega}$ for each grid cell. First, we find the smallest rectangle that covers the city of Cambridge and then discretize the map into $u \times v$ grid cells. 
\begin{figure*}[t!]
\centering
  \includegraphics[width=\textwidth]{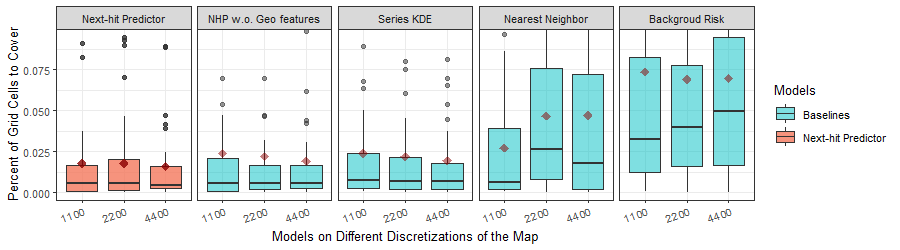}
\caption{The normalized ranks of true cells on the test set. The red dot represents the average}\label{fig:box}    
\end{figure*}

To obtain the geographic features, we use the land use datasets from the GIS system available at Boston public data portal\footnote{\url{www.cambridge.gov}}. The land use datasets contain a set of shape files for buildings and constructions of different types.  We extract features for different types of land use, including the public open space, single-family residential building, two-family residential building, commercial building, apartment building, transportation, and etc. 

We compute the area and the number of buildings of each subtype in each grid cell. When counting the number of buildings, since one building could span multiple grid cells, we count the proportion of the building that overlaps with the grid cell to avoid repetitive counting. 
We extract the area and count information for all subtypes of land use. In addition, we extract the asset value for residential buildings, since the wealth of the neighborhood is also a factor in when a criminal chooses a crime location. Some offenders are more likely to be attracted to the wealthy neighborhood while others might be turned away with the high-security level at in wealthy neighborhoods.  
For the MBTA feature, we compute the distance to the closest MBTA station for each grid cell, to represent the ease to get away from the crime scene.


\section{Experiments}\label{sec:experiments}
We used data from 4916 housebreaks in Cambridge, MA between 1997 and 2006
recorded by the Crime Analysis Unit of the Cambridge Police Department. Among the 4916 housebreaks, 682 are serial crimes from 55 crime series collected over the same period of time that were
curated and hand-labeled by crime analysts. For every crime series $p \in \{1,\cdots,P\}$, we reserved the last crime as testing, i.e., $\mathcal{C}_\text{test} = \{i |p_i \in \mathcal{P},  t_i = \max_{k \in \mathcal{C}_p} t_k\}$, and the remaining crimes for training and validation.

\noindent\textbf{Baselines}
To compare with the proposed model, we constructed four baseline models. All of the four models are based on the discretized grid map we mentioned in the second section. For the first baseline, we would like to see the effect of the geographic features in the NHP model. So we replace the numerator in the triggered risk in formula (2) with 1 while everything else in NHP remains the same. The first baseline also serves as an ablations study. Then we investigate the contribution of the information provided by prior offenses in the series. So we design two baseline models using only the prior offenses in the series. One uses kernel density estimation. The other uses 
 the nearest neighbor that computes the sum of the distance between each cell to all of the previous crimes for each cell: the shorter the distance, the higher the risk score.  For the last baseline, we would like to know the effect of the background risk. So we compute the risk by kernel density estimation based on $T$ days of the historical crime events before the prediction time. The parameters in the kernels in the above baselines and $T$ are tuned via cross-validation. 



\noindent\textbf{Experiment Setup and Evaluation} We discretize the map into a set of grid cells $\mathcal{L}$. Then for a test case in crime series $p$ at time $t$, a risk map is generated for a model. Grid cells in the map are ranked according to the risk, from the highest and lowest. We then record the rank of the true location, denoted as $\text{rank}(l^{\langle p,t\rangle}_*)$. We try different discretizations of the map into 1100, 2200, and 4400 cells to achieve different resolutions of the map. For a fair comparison, we report the \emph{normalized rank}, $\frac{\text{rank}(l^{\langle p,t\rangle}_*)}{|\mathcal{L}|}$, which represents the percentage of the grid cells one needs to cover in order to find the true crime. 
The smaller the normalized rank is, the better the prediction.
We plot the normalized ranks for the 55 test crimes in Figure \ref{fig:box}. We found that the proposed model demonstrated consistently good performance across different maps with the best one (smallest mean and median) being $\mathcal{L}=4400$, where each grid cell is a 70m by 70m square.

NHP without geographic features and series KDE achieve the closest competitive performance compared to NHP and the background risk only performs very poorly, which indicates that the behavior of each criminal is largely determined by his/her previous offenses instead of the general criminal cohort. Meanwhile, geographic features to improve the predictive accuracy. Compared to the self-exciting point process models, KDE does not consider the temporal decaying effect, thus achieving a slightly worse performance.

We remark that Next Hit Predictor has several advantages over the competing models: (i) the parameters are trained rigorously via a global objective; (ii) the triggering kernel function considers the spatial-temporal features and also the geographic features; and (iii) the model considers the both the general preference of the entire crime cohort and also individual preferences of criminals. 

\section{Conclusion}
We proposed a novel method, Next Hit Predictor (NHP), to predict the location of the next serial crime, given the prior offenses in the series.  NHP adopts the framework of self-exciting point processes created for modeling earthquakes, to characterize the correlations between crimes committed by the same criminal. In this way, NHP does not need to model the latent preference of an unknown criminal but directly represents the individual preference using the triggering kernels. We designed a new kernel that considers spatial, temporal and geographic correlations between crimes. To rigorously train the model, we formulated a convex learning objective that guarantees global optimum.
 
 While crime prediction has been studied extensively for predictive policing, serial crime prediction has not received much attention despite the importance of the problem. We believe our model has a great potential in this field and can help the police better allocate resources when solving and preventing serial crimes.
 \bibliographystyle{aaai}
 \bibliography{crimeseries}

\end{document}